# Blue shift of yellow luminescence band in self-ion-implanted n-GaN nanowire

S. Dhara,[a)] A. Datta,[b)] C. T. Wu, Z. H. Lan, K. H. Chen,[*] Y. L. Wang

Institute of Atomic and Molecular Sciences, Academia Sinica, Taipei, Taiwan

C. W. Hsu, L. C. Chen

Center for Condensed Matter Sciences, National Taiwan University, Taipei, Taiwan

H. M. Lin, C. C. Chen

Department of Chemistry, National Taiwan Normal University, Taipei, Taiwan

Y. F. Chen

Department of Physics, National Taiwan University, Taipei, Taiwan

Abstract

Optical photoluminescence studies are performed in self-ion ($Ga^+$)-implanted nominally doped n-GaN nanowires. A 50-keV $Ga^+$ focused ion beam (FIB) in the fluence range of $1 \times 10^{14}$ - $2 \times 10^{16}$ ions $cm^{-2}$ is used for the irradiation process. A blueshift is observed for the yellow luminescence (YL) band with increasing fluence. Donor-acceptor pair (DAP) model with emission involving shallow donor introduced by point-defect clusters related to nitrogen vacancies and probable deep acceptor created by gallium interstitial clusters is made responsible for the shift. High temperature annealing in nitrogen ambient restores the peak position of YL band by removing nitrogen vacancies.

[*] Corresponding Author : chenkh@po.iams.sinica.edu.tw

[a)] The author is presently on leave from Materials Science Division, Indira Gandhi Centre for Atomic Research, Kalpakkam–603102, India

[b)] Now at Netaji Nagar Day College,170/436 N.S.C Bose Road, Kolkata-700092, WB, India

Extensive studies[1-8] are performed in understanding various defects present in n- and p-GaN film as defects play an important role in optical and electrical properties of semiconductors for the device fabrication. In the photoluminescence (PL) studies, yellow luminescence (YL) band around 2.2 eV (band width ~1 eV) is one of the most well discussed defect bands present in GaN film with either native point-defects[3-5] or point-defects nucleating at extended defects like dislocations[6] as the origin. However, there seems to be an agreement that transitions from the conduction band or a shallow donor to a deep acceptor are responsible for this band. Properties of YL band, e.g., metastablity[7] and blueshift under hydrostatic pressure,[4,8] have also been discussed for the case of GaN samples. Though effects of ion irradiation, an indispensable tool for modern device fabrication, on GaN film have been reported[9] in a great deal but optical properties of ion irradiated GaN film are completely missing in the literature. Moreover, there is hardly any report on the study of defects in GaN nanowire, which has opened up a new avenue for the application of one dimensional (1-D) optoelectronic nanodevice.[10] In our earlier study, we have shown defect mobility is higher in 1-D system than that of higher dimensional systems, leading to enhanced dynamic annealing (defect annihilation by ion beam) in the $Ga^+$-implanted GaN nanowires.[11]

In this letter, we discuss the properties of YL band in case of self-ion ($Ga^+$)-implanted GaN nanowire. The nanowires, grown by catalyst assisted atmospheric pressure chemical vapor deposition (APCVD) technique, were irradiated with 50 keV $Ga^+$ using a focused ion beam (FIB). Self-ion implantation, a chemically clean process, was used to remove ambiguities of compound formation as reported in the earlier report.[9]

GaN nanowires (~25-100 nm) were grown on c-Si substrate coated with Au catalyst, using Ga as source material and $NH_3$ (10 sccm) as reactant gas in a tubular furnace (substrate temperature 900°C) by APCVD technique. The grown nanowire ensemble was found to be nominally doped n-type from our Hall measurements with carrier concentration $\sim 3 \times 10^{13}$ cm$^{-3}$. Self-ion implantation on these GaN nanowires was studied using a Ga$^+$ FIB at 50-keV with beam current of ~1.3 nA in the fluence range of $1 \times 10^{14}$ - $2 \times 10^{16}$ ions cm$^{-2}$. For 50 keV Ga$^+$ in GaN film, SRIM code calculation shows that the energy dissipation is mainly through nuclear energy loss $(dE/dx)_n \sim 2$ keV/nm. PL measurements were performed using He-Cd laser tuned to 325 nm with an output power of ~10 mW at room temperature (RT). The emission signal was collected by a SPEX 0.85-m double spectrometer and detected by a lock-in-amplifier.

Structural study of one of the pristine nanowires with high resolution transmission electron microscopy (HRTEM) (Fig. 1) confirmed the growth of wurtzite (hexagonal) GaN (h-GaN) with zone axis lying along [001] direction as calculated from the corresponding selected area electron diffraction (SAED) pattern in the inset. Crystalline order is maintained all along the nanowire as observed from lattice imaging (Fig. 1).

PL study of the pristine and the irradiated samples in the fluence range of $1 \times 10^{14}$ - $2 \times 10^{16}$ ions cm$^{-2}$ at RT is shown (Fig. 2). Direct band-to-band transition around 3.42 eV shows the signature of wurtzite phase present in the pristine sample as also confirmed from the HRTEM lattice imaging and the SAED analysis (Fig. 1). The intensity of the band-to-band transition peak (Fig. 2), is observed to deteriorate with increasing fluence. This may be due to the increase in disorder introduced during the irradiation process with increasing fluence. A band edge transition peak around 3.1 eV, designated as longitudinal optical (LO) phonon replica mode of the donor-acceptor pair (DAP) for cubic phase,[2,10,12] is observed (Fig. 2) in the pristine sample. The cubic phase is reported to be stable as stacking faults embedded in the hexagonal structure of GaN nanowire.[10] Introduction of defects by ion irradiation process, however, stabilizes the cubic phase as the ratio of intensities of LO phonon replica mode of DAP (corresponding to cubic phase) to band-to-band transition peak (corresponding to hexagonal phase) is observed to increase with increasing fluence ratio (inset of Fig. 2). The zinc-blend symmetry of the cubic phase is reported to be stable at vacancy like point-defects in the GaN using *ab initio* molecular dynamic calculations.[13] We have also observed a peak around ~2.2 eV for the pristine sample (Fig. 2) which can be identified as YL band of GaN nanowire. The peak of YL band shows a large blue shift with increasing fluence (Fig. 2) with the peak position at ~2.8 eV for the sample irradiated at a fluence of $2 \times 10^{16}$ ions cm$^{-2}$. The BL band (~2.7-2.9 eV)[14,15] is observed for nominally (unintentionally) doped and Si-doped n-GaN. BL band in n-GaN was initially attributed to homogeneously distributed point defects, which was extended to DAP model by Reschikov *et al.*[14] In their model, transition between conduction band as shallow donor and Ga vacancy ($V_{Ga}$) or its complexes, created during the accumulation of dislocation in unoptimized buffer layer, as deep acceptor level were proposed. BL band in n-GaN (nominally doped, and Si doped) and in p-GaN (heavily doped with Mg) is argued by Kaufman *et al.*[15] to have the same origin and differ only in concentration of the related defects. Blue luminescence (BL) band at 2.8 eV is reported for heavily doped p-GaN film,[15,16] with transitions from shallow donors (originating from $V_N$ complexes) to deep Mg acceptor levels following the same DAP model.

The defect structures in the irradiated samples were analyzed using HRTEM images [Figs. 3(a)-(d)] and details can be seen in our earlier report.[11]. Accumulation of point-defects and role of dynamic annealing which is efficient only at optimum fluence are discussed for the microstructural evolution in these samples. The defect structures in the irradiated samples suggest agglomeration of point-defect clusters as the major component of disorder at high fluences. Enhanced dynamic annealing,[11] with high diffusivity of mobile point-defects in the nanowires, might have prohibited the growth of extended defects (reported for the heavy ion irradiated epi-GaN)[9] in the irradiated samples. These point-defect clusters might be containing $V_N$ as major component, as the ion irradiation in GaN can release nitrogen as material dissociation takes place in the damage process.[17] Presence of vacancy like point-defects is also discussed earlier from the point of view of stabilizing cubic phase with increasing fluence (inset of Fig. 2). Analytical electron microscopic (AEM) analysis showed [inscribed in Fig. 3(a)-(d)] increasing nitrogen deficiencies in the irradiated samples with an increasing ion fluence. Large energy deposition during the 50-keV self-ion ($Ga^+$) implantation in GaN nanowire favor the formation and accumulation of $V_N$, as the nitrogen vacancy formation energy is only 4 eV/$V_N$.[18] These $V_N$ clusters can play a role of shallow donor states,[18] as also reported for Mg-doped GaN.[16] Simultaneously, accumulation of Ga interstitials ($Ga_I$) in forming a cluster is also likely in the self-ion ($Ga^+$) implantation process, as interstitial formation energy (~10 eV/$Ga_I$)[18] involved is also provided in this energetic process. Formation of $Ga_I$, stable close to RT, is reported by purely electronic energy loss process of 2.5 MeV electron irradiation in epi-GaN.[19] This argument is purely from the energetics involved in our nuclear energy loss dominated process and $Ga_I$ formation energy. At present, there is no literature available for the optical identification of $Ga_I$ clusters. Molecular dynamic (a*b initio*) calculation showed that the relaxed $Ga_I$ (with tetrahedral symmetry) in GaN creates a deep trap at 0.7 eV above the valence band and act as deep acceptor.[4] The appearance of a broad BL band may be argued from the DAP model with clusters of $V_N$ and $Ga_I$ giving rise to the shallow donor and deep acceptor states, respectively, for the electronic transition. It seems that the $V_N$ complex and $Ga_I$ clusters must have produced multiple states or even a broad distribution of states which lead to progressive change in the peak position towards higher energy of an intrinsically broad band.

In order to confirm the role of $V_N$, the samples were annealed in a two-step process, 15 minutes at 650 $^0$C and 2 minutes at 1000 $^0$C in $N_2$ ambient. The YL band, measured at RT, for the post-annealed irradiated samples is observed (Fig. 4) to shift back to ~ 2.2 eV, which is the original position of YL band of the pristine sample (Fig. 2). With clusters of $V_N$ being removed in the $N_2$ ambient annealing process and clusters of $Ga_I$ being unstable at high temperature, the YL band is expected to restore back to its original position. The intensities of the YL band of post–annealed samples are observed (Fig. 4) to increase with increasing fluence containing larger defects during growth. The intensity fluctuations of YL band (Fig. 2) with increasing fluence in the as-irradiated samples may be due to complex defect formation during the irradiation process and removed in the annealing process.

In conclusion, the yellow luminescence (YL) band in nominally doped n-GaN nanowire shows a blueshift with accumulation of nitrogen vacancies during the, chemically clean, self-ion irradiation process with a blue luminescence band observed at ~ 2.8 eV for the sample primarily containing large nitrogen vacancy related point-defect clusters at the fluence of amorphization. Transitions involving shallow donor related to $V_N$ clusters and probable deep acceptor linked to $Ga_I$ clusters in the energetic irradiation process might be responsible for the blueshift.

We acknowledge National Science Council and Ministry of Education in Taiwan for financial assistance.

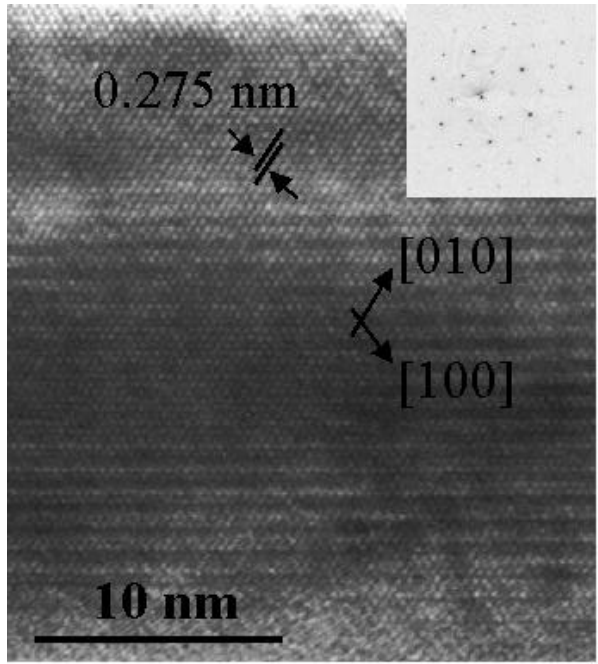

Fig. 1. HRTEM image of one of the pristine GaN nanowires. The lattice spacing of 0.275 nm corresponds to distance between two (100) planes of h-GaN. SAED pattern, in the inset, showing the formation of wurzite phase with zone axis lying in [001] direction.

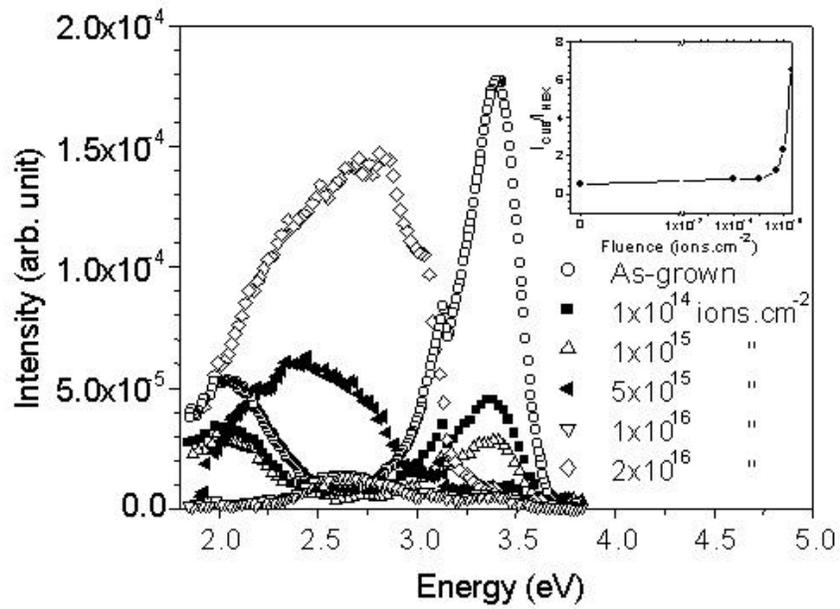

Fig. 2. Room temperature PL study of the pristine and irradiated GaN nanowires in the fluence range of $1\times10^{14}$-$2\times10^{16}$ ions cm$^{-2}$ showing a blue shift of the YL band ~ 2.2 eV with increasing fluences. Inset shows increasing ratio of intensities corresponding to cubic and hexagonal phases with increasing fluence. The line is guide to an eye.

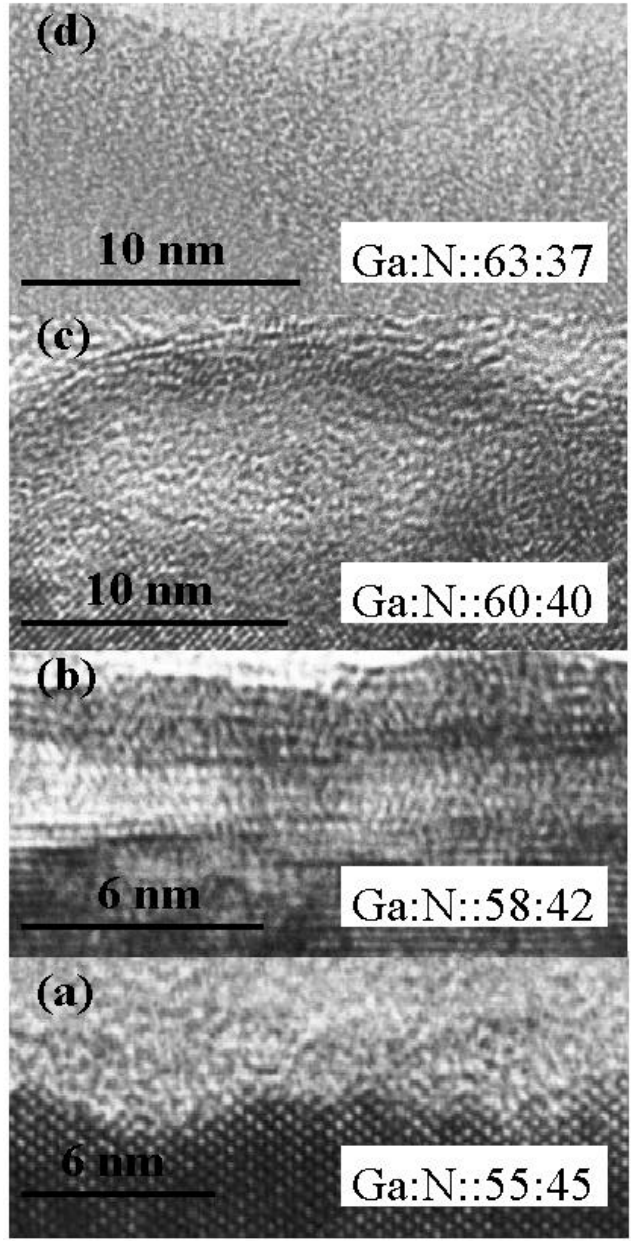

Fig. 3. HRTEM analysis of the defect structures of irradiated GaN nanowires for the fluences of (a) $1\times10^{15}$ ions cm$^{-2}$ (b) $5\times10^{15}$ ions cm$^{-2}$ (c) $1\times10^{16}$ ions cm$^{-2}$ and (c) $2\times10^{16}$ ions cm$^{-2}$. The AEM compositional analysis is inscribed for the respective samples showing increasing nitrogen deficiencies with increasing fluence.

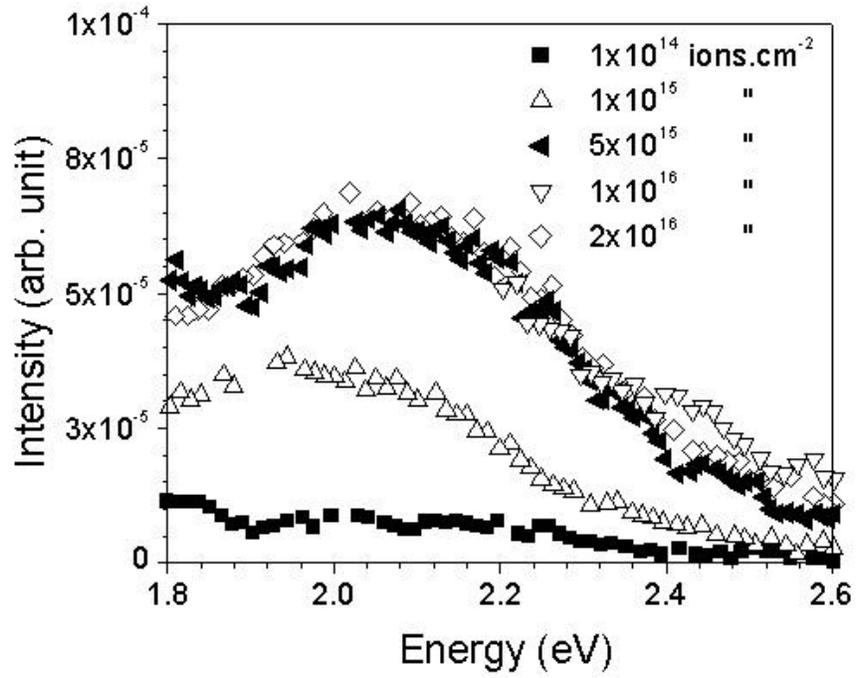

Fig. 4. Room temperature PL study of the post-annealed samples of GaN nanowires irradiated in the fluence range of $1\times10^{14}$ - $2\times10^{16}$ ions cm$^{-2}$ showing restoration of YL band ~ 2.2 eV in the two-step annealing treatment in $N_2$ ambient.